\documentclass[a4,11pt]{cip-v3-cls-submit}
\usepackage{booktabs}
\usepackage{cite}
\usepackage{mathptmx}
\DeclareSymbolFont{epsilon}{OML}{cmm}{m}{it}
\DeclareMathSymbol{\epsilon}{\mathord}{epsilon}{"0F}
\usepackage{hyperref}
\hypersetup{
	colorlinks=false,
	pdfborder={0 0 0}
}
\usepackage{graphics,graphicx,subfigure,wrapfig,epstopdf}
\usepackage{wasysym,amsmath,amsxtra,amssymb,latexsym,amscd,color,cite,bm,array,multirow}
\usepackage[mathscr]{eucal}
\def\authors#1{\author{\begin{flushleft}{#1}\end{flushleft}}}
\def\authord#1#2{\textbf{\indent{#1}$^{#2}$}}
\def\addressed#1#2{\\[1mm]\textit{$\!\!\!^{#1}$\indent#2}}
\def\CorrEmail#1{\\[4mm]
	\textit{E-mail:}~$^\dag${#1}}

\def\Keywords#1{$\qquad$\\[-.35cm] \textnormal{Keywords:~{#1}}.} 

\def\and{\textbf{and} }

\begin{document}

	\title{Spatially heterogeneous noise restructures flocking into geometry-locked and vortex states}
	\authors{
	\authord{Ankush Semwal}{1}, \authord{Mahak Poonia}{2}, \authord{Pintu Patra}{1,\dagger}
	\newline
	\addressed{1}{Indian Institute of Technology Kharagpur, West Bengal, India}
	\addressed{2}{Institut für Physik, Johannes Gutenberg-Universität Mainz, Mainz,
Germany}
\CorrEmail{pintupatra@phy.iitkgp.ac.in}
}
\maketitle
	\markboth{Spatially heterogeneous noise restructures flocking $\ldots$}{Semwal \textit{et al.}}

\begin{abstract}
Spatially heterogeneous environments continually challenge the ability of active matter to sustain coherent collective motion. Understanding how collective motion remains robust under changing environments is central to both the functioning of biological systems and the design of smart active matter. Here, we extend the Vicsek model to include a circular non-noisy region surrounded by a noisy environment $-$ a configuration in which the noise difference sets up a contrast in local directional order between the two regions. We find that, as the surrounding noise is increased, the system passes through three distinct dynamical regimes: (i) conventional global flocking at low noise; (ii) geometry-locked motion, aligned with simulation boundaries, at intermediate noise; and (iii) vortical motion within the non-noisy region at high noise. Extending the environment to multiple non-noisy regions, we find that the geometry-locked regime can develop a directional coupling, while the vortex mode leads to antiferromagnetic order between the regions. Taken together, our results demonstrate that the spatial modulation of order and disorder offers a powerful and generic strategy for steering active matter, aligning with recent experimental observations of active particles in patterned landscapes.
\end{abstract}



\newpage
\section{Introduction}

Living active matter, ranging from groups of animals and microorganisms to cytoskeletal filaments, displays remarkable examples of collective motion. In natural settings, living active matter often navigates through heterogeneous environments \cite{Bechinger2016}, such as patchy food landscapes \cite{Jacucci2024}, the presence of obstacles \cite{Morin2017, Chepizhko_Oleksandr2013, Vahabli2023} or geometric constrictions \cite{Volpe2011, Wioland_2016, Theillard2017, Modica2023}, changing weather conditions \cite{breuner2013environment, van2018continental, Couzin2018}, or interactions with other individuals \cite{Couzin2005} that influence their collective behavior. In contrast to homogeneous landscapes, such spatial variations can give rise to distinct subsystems that favor different degrees of local order, resulting in a direct competition between locally and globally ordered configurations \cite{Yang2025, Morin2017, Soker2021}. Experimentally, heterogeneous landscapes containing specific regions with differential behavior have recently emerged as a powerful route for dynamically tuning and programming active matter \cite{Souslov2017}. Several studies have demonstrated that active systems can be manipulated by applying non-uniform spatial fields \cite{Zhang2022, Han2020multivortex, Arlt2019}, including geometrical confinement \cite{Wioland_Hugo2013, Qu2021, Wu2017, Lushi2014, Liu2021}, micro-indented surfaces \cite{Galajda2007}, or local optical \cite{Arlt2018, Jahanshahi2020, Massana-Cid2022} and acoustic activation \cite{Takatori2016, Han2023}. These studies have shown that spatial heterogeneity leads to non-uniform accumulation \cite{Xu2024}, trapping \cite{Chepizhko2013}, and correlated transport of active particles \cite{bricard2013emergence, Wioland_2016, Xu2024}. However, despite numerous reported control strategies for active matter, the physical principles governing the directional control of such emergent systems, as well as their robustness to parameter variations, remain poorly understood.

 The persistent gap in understanding these physical principles has prompted a growing body of theoretical studies aimed at both understanding experimentally observed behaviors and suggesting new control strategies \cite{Katuri2024, liu2021activity, Liu2021, Chepizhko_Oleksandr2013}. For instance, a recent work established directed transport by optimally controlling a localized spotlight region via a reinforcement learning algorithm \cite{falk2021learning}. Furthermore, models consisting of spatial regions with zero noise have been exploited to achieve directional steering \cite{SpatialNoise}, sorting \cite{Huang_jun2024}, migration \cite{Huang2024}, and trapping of active particles \cite{Huang2025}. However, the physics driving the emergence of these collective states remains unclear. Furthermore, how these states scale or correlate in larger systems with multiple regions remains unexplored.
 
To address these open questions, here we investigate the Vicsek model in the presence of spatially heterogeneous noise, which serves as the primary control parameter driving the system's order–disorder transition. Specifically, we extend the model by introducing a zero-noise circular domain embedded within a noisy surrounding region, a geometry selected to preserve rotational symmetry and ensure the absence of any inherent directional bias. Through numerical simulations, we uncover three distinct dynamical regimes that emerge as a function of the noise contrast between these two environments. Beyond conventional Vicsek-like flocking, we observe geometry-locked collective motion at intermediate surrounding noise, and confined circular motion at high surrounding noise. Our results demonstrate how spatial heterogeneity restructures flocking into novel collective states. These states can be easily selected and correlated in space by tuning model parameters.

\section{RESULTS}

\section{Model}

\begin{figure}
    \centering
    \includegraphics[width=1\linewidth]{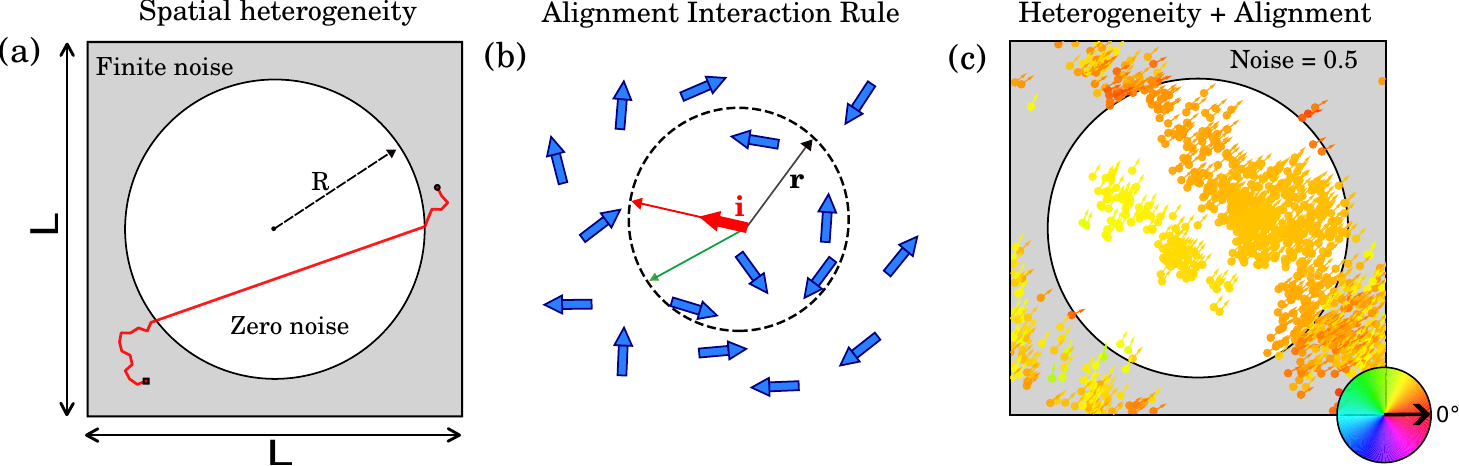}
    \caption{\textbf{Model setup and representative simulation snapshot.} (a) Illustration of the model setup. A non-noisy circular region of radius $R$ is embedded at the center of a noisy surrounding domain. The solid line represents a simulated trajectory of an individual particle. (b) Schematic of the alignment interaction rule. At every time step, a particle's orientation is updated based on the average orientation of all particles in its neighborhood (blue arrows within the interaction radius $r$). The red arrow indicates the orientation of the focal particle $i$, while the green arrow represents the average direction of neighbors and the future direction of the focus particle. (c) Simulation snapshot at $t = 500$ for a system of $N = 1000$ particles. External noise strength $\eta = 0.5$, particle speed $v = 0.5$, interaction radius $r = 1$, non-noisy region radius $R = 8$, and system size $L = 20$.}
    \label{fig:fig1}
\end{figure}

Our simulation setup consists of a spatially heterogeneous version of the Vicsek model, in which individuals move in two-dimensional space containing a circular non-noisy domain embedded at the center of a tunable noisy surrounding \cite{SpatialNoise, Huang_jun2024, Huang2024, Huang2025, khan_2026}. The model is depicted in Figure~\ref{fig:fig1}(a). Within this setup, an isolated particle shows linear motion along its initial direction in the non-noisy region and non-linear motion in the surrounding region due to directional randomization caused by finite noise. The direction of particles changes upon interacting with other particles following Vicsek interaction rules \cite{vicsek1995novel}. At every time step, the direction of a particle is updated with the mean orientation of all the particles within their interaction radius, regardless of their location. Further, a finite random noise is added to the average direction.  The dynamics of the particles in our model can be summarized through the following equations, 

 \begin{eqnarray}
{r}_i(t+1) &=& {r}_i(t) + v(\theta_i)\,\Delta t, \\
\theta_i(t+1) &=& \langle \theta(t) \rangle_{|{r}_{ij}| < r} + \xi (t) \times
\begin{cases}
0, & d<R \text{ (inside)}, \\
\eta, & d>R \text{ (outside)},
\end{cases}
\end{eqnarray}
where
\[\langle \theta(t) \rangle_{|{r}_{ij}| < r}
= \tan^{-1}\!\left(
\frac{\sum_{j} \sin \theta_j(t)}{\sum_{j} \cos \theta_j(t)}
\right)\].

is the average direction all particles $j$ within the interaction radius $r=1$ of the focus particle $i$ (as represented in Figure~\ref{fig:fig1}(b)). Here, $\eta$ is the strength of the orientational noise, $\xi $ is a uniformly distributed random variable from $(-1/2, 1/2)$, $R$ is the radius of the non-noisy circular domain, and $d = \sqrt{(x_i - L/2)^2 + (y_i - L/2)^2}$ is the Euclidean distance of the particle from the center of the simulation box (of dimensions $L \times L$). Throughout the study, we have set $v=0.5$ and $\Delta t=1$ for the sake of simplicity. Further, the particles are subjected to periodic boundary conditions. For this study, we have considered surrounding noise ranging from $0$ to $5$. Figure~\ref{fig:fig1}(c) shows a typical simulation snapshot where noise between the two domains is comparable ($\eta= 0.5$). In this example, we observe a strong directional alignment between the particles throughout the space.

\section{Spatially heterogeneity restructures flocking motion into new emergent modes}

\begin{figure*}[t]
    \centering
    \includegraphics[width=1\linewidth]{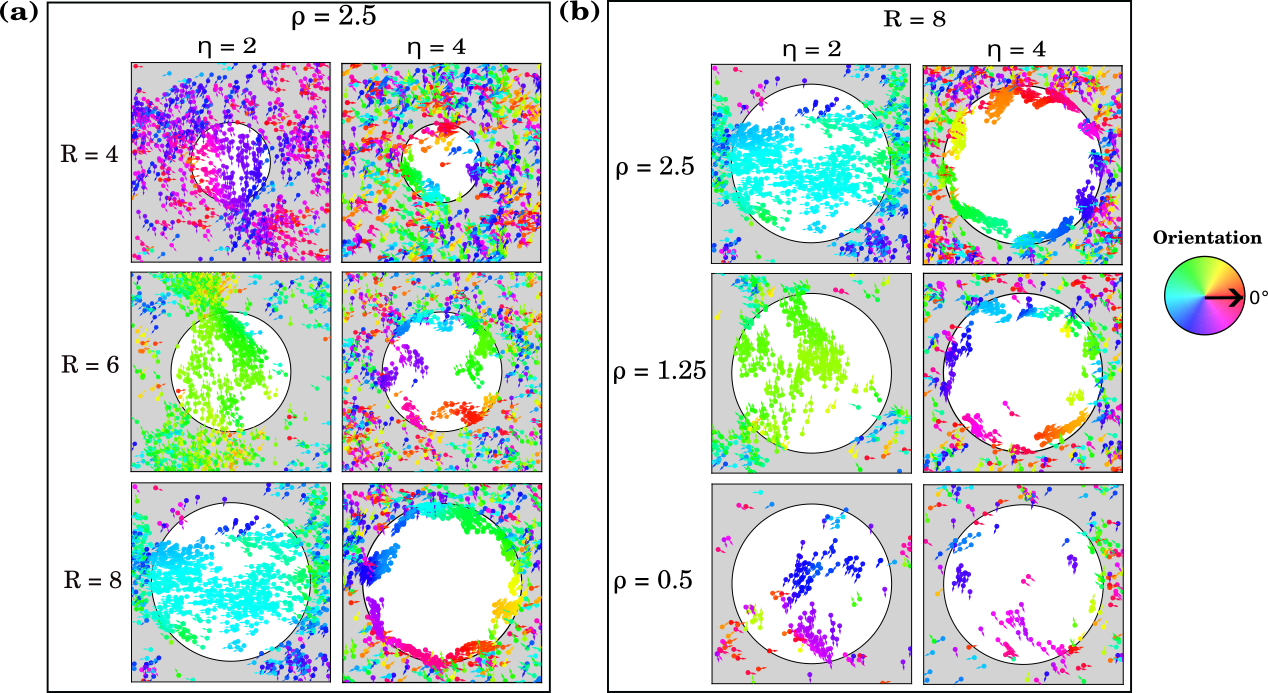}
    \caption{\textbf{Dependence of collective motion on model parameters.} (a) Flocking patterns at number density $\rho = 2.5 (N=1000)$ for non-noisy region radii $R = 4$, $6$, and $8$ (top to bottom rows) and surrounding external noise strengths $\eta = 2$ and $4$ (left and right columns, respectively). (b) Flocking patterns for particle densities $\rho = 2.5$, $1.25$, and $0.5$ (top to bottom rows) at external noise strengths $\eta = 2$ (left column) and $4$ (right column), with the non-noisy region radius fixed at $R = 8$. Unless otherwise stated, all parameters are fixed at the values given in Table~\ref{tab:parameters}.}
    
    \label{fig:fig2}
\end{figure*}

We first examine how the surrounding noise and the size of the non-noisy region shape the flocking patterns. Figure~\ref{fig:fig2}(a) presents the collective patterns at a high particle density ($\rho=2.5$) for different sizes of the non-noisy region with radius ranging from $R=4$ to $R=8$. At moderate noise values ($\eta=2$), increasing the size of the non-noisy region progressively strengthens the flocking motion. For a small region ($R=4$), the individuals show weak alignment within the non-noisy region. As the region size increases ($R=6 \text{ and } 8$), the alignment progressively strengthens, yielding a high global directional alignment along the simulation box (see Supplementary Movie 1). For a high value of outside noise ($\eta=4$, Figure~\ref{fig:fig2}(a), right column), the particles outside remain disordered, while the particles inside favor a highly ordered state due to zero noise. The outcome of this conflict is determined by the size of the non-noisy region. For a small region ($R=4$), the internal order is continuously destroyed as the flock moves outside. As the region size is increased ($R=6$), the flocks within the domain preserve their local alignment by moving along the inner interface (inside the non-noisy region). With a further increase ($R=8$), a vortex state emerges, which shows collective vortical motion along the inner interfaces  between the two regions (see Supplementary Movie 2 and Supplementary Fig.~S1 for different patterns depending on region size and noise strength).

We next examine the role of particle number density for a fixed non-noisy region radius ($R=8$). Figure~\ref{fig:fig2}(b) shows long-time snapshots for particle densities, $\rho=2.5$, $1.25$, and $0.5$ at noise strengths $\eta=2$ and $\eta=4$. For intermediate noise ($\eta=2$), lowering the density of the particles progressively weakens the directional order. The system exhibits coherent directional flocking at high density ($\rho=2.5$), whose strength weakens at low density ($\rho=1.25$). Upon further reduction of density to $\rho=0.5$, coherent flocking is lost altogether. At high noise and high density ($\eta=4$, $\rho=2.5$), the circular motion along the inner interface becomes more prominent. As the density decreases ($\rho=1.25$), its tendency to move along the interface becomes less pronounced; and with a further reduction ($\rho=0.5$), the particles form small flocks that move randomly (see Supplementary Movie 3). Taken together, these observations show that the surrounding noise dictates the nature of the collective motion inside the non-noisy region, leading to different modes of collective motion. To characterize these emergent states, new metrics and order parameters must be defined for the observed patterns.

\begin{table}[t]
\centering
\footnotesize 
\fbox{%
\begin{minipage}{0.7\linewidth}
\caption{Default value of the parameters.}
\label{tab:parameters}
\begin{tabular}{p{3.2cm} p{2.0cm} p{5.5cm}}
\toprule
\textbf{Parameter} & \textbf{Default value} & \textbf{Definition} \\
\midrule
Particle density ($\rho$) & 2.5 &
Number of particles per unit area ($N/L^2$) \\
dt & 1 & Time step \\
$T$ & 500 & Total simulation time \\
$v_{0}$ & 0.5 & Speed of the particles \\
$r$ & 1 & Alignment interaction radius \\
$L$ & 20 & Length of the simulation box \\
\bottomrule
\end{tabular}

\end{minipage}}
\end{table}

\section{Quantitative characterization of the emergent collective states}

\begin{figure}[t]
    \includegraphics[width=1\linewidth]{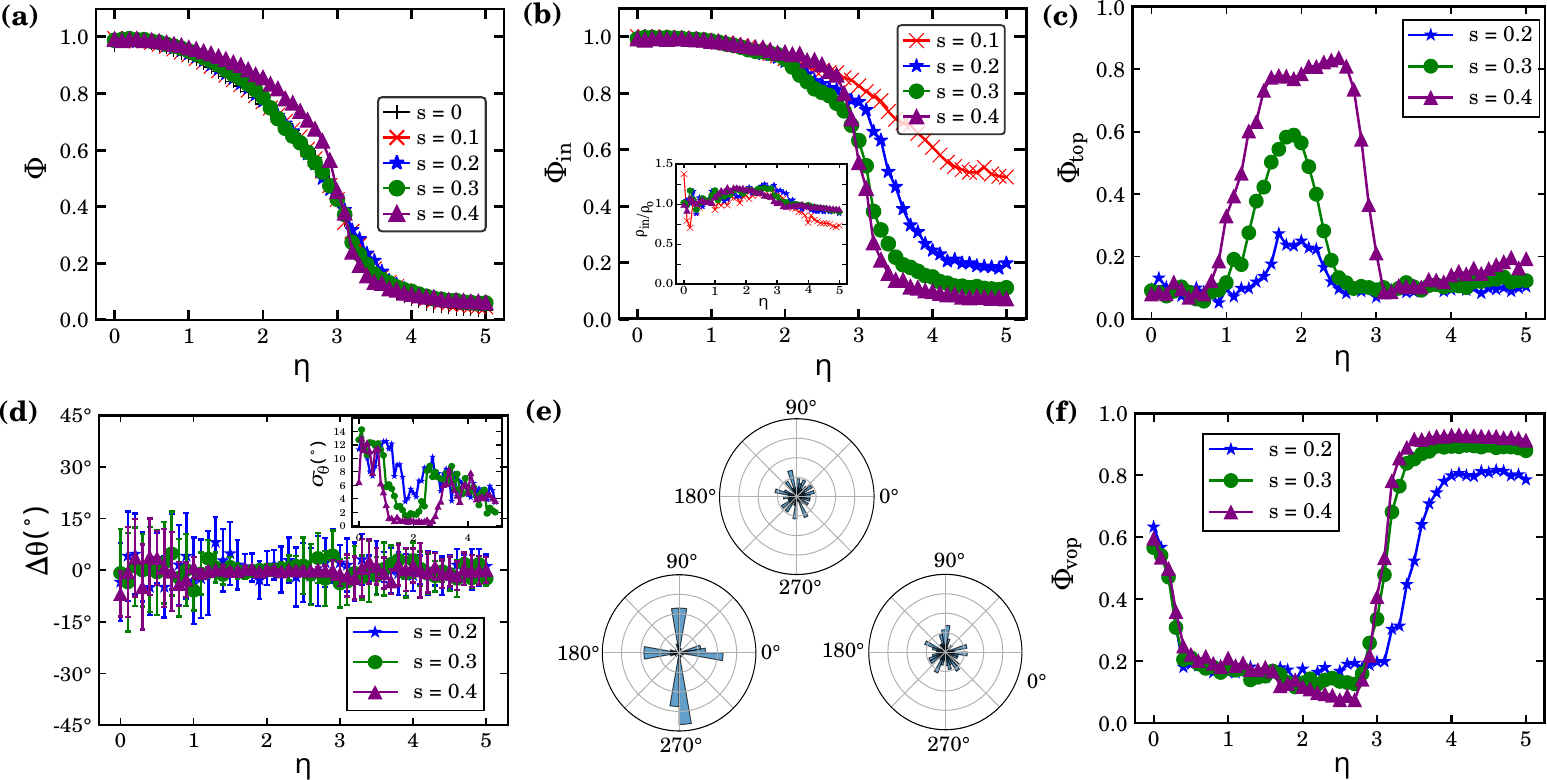}
    \caption{\textbf{Variation of order parameters with surrounding noise strength.} (a, b) Global ($\Phi$) and local ($\Phi_{\mathrm{in}}$) Vicsek order parameters as a function of surrounding noise ($\eta$) for different non-noisy region sizes ($s$). The inset in (b) shows the normalized particle density inside the non-noisy region, $\rho_{\mathrm{in}}/\rho$, versus $\eta$. (c) Tetratic order parameter ($\Phi_{\mathrm{top}}$) as a function of $\eta$ for various values of $s$. (d) Angle of tetratic order within the region versus $\eta$; the inset displays the standard deviation $\sigma_{\theta}$ of the angle as a function of $\eta$. (e) Distributions of particle orientation angles within the non-noisy region for $\eta = 0.5$ (top), $\eta = 2.0$ (bottom left), and $\eta = 5.0$ (bottom right), obtained from $100$ simulations runs. (f) Vortex order parameter ($\Phi_{\mathrm{vop}}$) as a function of $\eta$ for varying $s$. Remaining parameters are provided in Table~\ref{tab:parameters}.
    }
     \label{fig:fig3}
\end{figure}

We begin by quantifying the Vicsek order parameter, namely, the magnitude of the mean normalized direction of all particles as a function of the surrounding noise (see Methods section for definition). Figure~\ref{fig:fig3}(a) displays the variation of the Vicsek order parameter, denoted by $\Phi$, with surrounding noise for several non-noisy region sizes, characterized by relative size, $s=R/L$, with respect to simulation domain size $L$. $\Phi$ monotonically decreases with noise similar to the Vicsek model for several non-noisy region sizes. The impact of the non-noisy region on the order parameter is minimal. Specifically, within the intermediate noise regime, a marginal increase correlates with the size of the non-noisy region. Figure~\ref{fig:fig3}(b) displays the variation of the Vicsek order parameter in the non-noisy region, denoted by $\Phi_{\mathrm{in}}$, for several non-noisy region sizes. In contrast to $\Phi$, the $\Phi_{\mathrm{in}}$ saturates at high noises to a finite value depending on the size of the non-noisy region. Surprisingly, for smaller non-noisy regions, a higher $\Phi_{\mathrm{in}}$ is observed. This non-trivial dependency stems from the emergence of circular motion for larger non-noisy regions (Fig.~\ref{fig:fig2}(a)), for which the Vicsek order parameter is low. This is due to the fact that for circular motion the individual direction vectors cancel one another, resulting in a smaller magnitude of the average direction vector of the enclosed particles. The order parameters of the surrounding subsystem, on the other hand, show trends similar to the Vicsek model (see Supplementary Figure S2(b)). Fig.~\ref{fig:fig3}(b) inset shows the relative number density of particles inside the non-noisy region. We observe a moderate accumulation of particles in the non-noisy region at intermediate noise levels only. This suggests that the observed deviation in the order parameter is primarily due to the emergent patterns and not due to the accumulation of particles. 

Physically, the linear flocking observed at intermediate noise arises because periodic boundary conditions continuously reinforce collective order along the shortest path for re-entering the non-noisy region. Because the non-noisy region is circular, it preserves rotational symmetry and exerts no directional bias. The alignment along the simulation axes, therefore, represents a explicit breaking of rotational symmetry, dictated entirely by the geometry of the setup. Hence, we refer to this emergent configuration as a geometry-locked state.

To quantify the strength of directional ordering in the geometry-locked state, we define the tetratic order parameter \cite{Hou2020} within the non-noisy region, denoted by $\Phi_{\mathrm{top}}$ (defined in the Methods section) (Fig.~\ref{fig:fig3}(c)). Simultaneously, to examine the deviation of the mean direction from the coordinate axes inside the non-noisy region, we compute the angle of the tetratic order, denoted by $\Delta \theta$ (defined in the Methods section) (shown in Fig.~\ref{fig:fig3}(d)). We find that for low surrounding noise ($\eta<1$), Vicsek-like order dominates; and consequently, $\Phi_{\mathrm{top}}$ is small, and $\Delta \theta$ fluctuates strongly. As the noise increases, $\Phi_{\mathrm{top}}$ increases progressively, and fluctuation in $\Delta \theta$ is suppressed, as is evident from the standard deviation $\sigma_{\theta}$ plot in Fig.~\ref{fig:fig3}(d) inset. As the noise is increased further, $\Phi_{\mathrm{top}}$ decreases to a low value due to the onset of local circular motion within the non-noisy region. Consequently, the variation of $\Delta \theta$ increases again. The transitions in different flocking modes are also captured in the histogram of the average orientation with the non-noisy region over multiple simulation runs (as shown in Fig.~\ref{fig:fig3}(e)). For low noise ($\eta= 0.5$, top distribution in Fig.~\ref{fig:fig3}(e)), the distribution is uniform over all directions, as in the Vicsek model. At intermediate noise ($\eta=2$, bottom left distribution in Fig.~\ref{fig:fig3}(e)), the histogram shows a bimodal distribution reflecting bias towards parallel or perpendicular collective alignment. A uniform distribution reappears at high noise ($\eta=5$, bottom right distribution in Fig.~\ref{fig:fig3}(e)), but this state is different from the Vicsek-like regime, as both $\Phi$ and $\Phi_{\mathrm{top}}$ are very small. This is a consequence of the emergence of circular motion in this regime, which we refer to as a vortex state.
 
To quantify the circular motion, we define the vortex order parameter $\Phi_{\mathrm{vop}}$ (defined in the Methods section). Fig.~\ref{fig:fig3}(f) plot its variation with $\eta$ for different non-noisy region sizes ($s=0.2,0.3,0.4$). $\Phi_{\mathrm{vop}}$ is close to zero for randomly oriented particles and approaches unity for vortically arranged ones. At $\eta<0.5$, $\Phi_{\mathrm{vop}}$ takes a finite value arising from spatial asymmetries, created by the formation of large flocks, in the particle distribution about the center of the domain. For intermediate noise, small flocks spread evenly across the space (see Supplementary Movie 4), and $\Phi_{\mathrm{vop}}$ decreases towards zero. For high noise, circular motion sets in, $\Phi_{\mathrm{vop}}$ increases, and eventually saturates to a finite value close to unity. 

In summary, the difference in collective ordering affinity between the two regions in our system introduces new modes that selects for geometry-locked linear flocking and collective vortical motion. While the geometry-locked flocking arises due to the periodic passage of the flock through the non-noisy region, the vortex motion is confined to the non-noisy regions, as the high outside noise immediately destroys any developed order.

\section{Phase diagram for distinct modes of motion}

\begin{figure*}[t]
    \centering
    \includegraphics[width=1\linewidth]{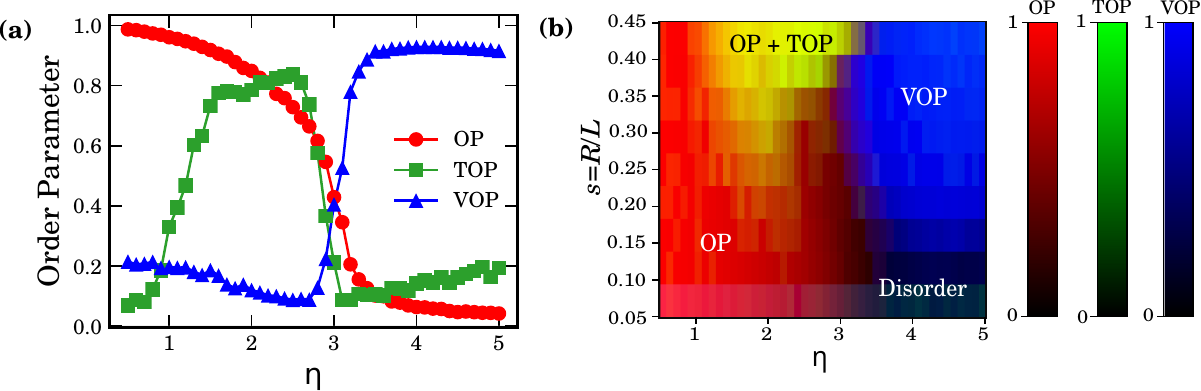}
    \caption{\textbf{Phase diagram showing different flocking modes.} (a) Evolution of order parameters as a function of external noise strength $\eta$ for a non-noisy region size $s = 0.4$. The Vicsek (OP), tetratic (TOP), and vortex (VOP) order parameters are represented by red circles, green squares, and blue triangles, respectively. (b) RGB phase diagram in $(s, \eta)$ space, where color intensity corresponds to the relative strength of the three collective states.}
    \label{fig:fig4}
\end{figure*}

To identify the three distinct dynamical modes observed in our system, we compare $\Phi$, $\Phi_{\mathrm{top}}$, and $\Phi_{\mathrm{vop}}$ as a function of $\eta$. Fig.~\ref{fig:fig4}(a) shows $\Phi$ (red circles), $\Phi_{\mathrm{top}}$ (green squares), and $\Phi_{\mathrm{vop}}$ (blue triangles) for different values of $\eta$. Since both $\Phi$ and $\Phi_{\mathrm{top}}$ reflect global linear flocking, their values are comparable at intermediate noise strength (approximately in the range of $1.5<\eta<3$), which makes it difficult to distinguish between these two states. However, the absence of tetratic order can be identified at low noises. In contrast, $\Phi_{\mathrm{vop}}$ clearly captures localized circular motion of flocks, a state for which both $\Phi$ and $\Phi_{\mathrm{top}}$ vanish. To visualize which mode of motion dominates in the parameter space, we plot an RGB color map by overlaying the three order parameters as intensity (255$\Phi$, 255$\Phi_{\mathrm{top}}$, 255$\Phi_{\mathrm{vop}}$) in Fig.~\ref{fig:fig4}(b). 

We observe that, for smaller non-noisy regions ($s < 0.15$), the system exhibits behaviour qualitatively similar to the Vicsek model. In this regime, the system transitions from an ordered to a disordered state without showing geometry-locked or vortex states. As the non-noisy region size increases ($s \geq 0.15$), the behavior deviates from the Vicsek-type order-disorder transition. For $s \geq 0.15$, $\Phi_{\mathrm{vop}}$ increases at high noise values ($\eta > 3.5$). A further increase in region size to $s \geq 0.3$ leads to the emergence of all three modes of motion in a sequential order with increasing noise. The geometry-locked and vortex states become more pronounced at larger sizes of non-noisy regions ($s > 0.30$). We also observe that in some cases, e.g., for $\eta \approx 3$, the system coexists or rapidly switches between the distinct modes observed. In conclusion, the phase diagram demonstrates that the spatial scale of the non-noisy region and the strength of the surrounding noise fundamentally shapes the emergent collective behavior of the system.

\section{Correlation of emergent states in multiple non-noisy regions}

\begin{figure*}[t]
   \centering
   \includegraphics[width=1\linewidth]{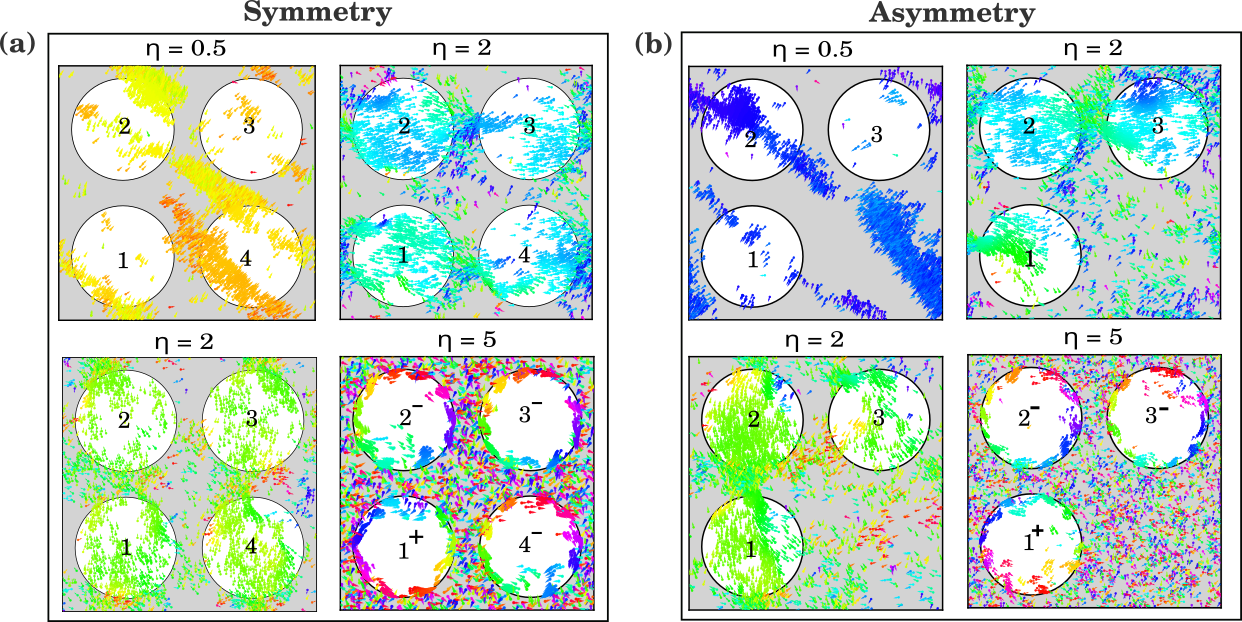}
    \caption{\textbf{Flocking patterns in symmetric and asymmetric arrangements of multiple non-noisy circular regions.} The non-noisy region size is fixed at $s = 0.4$, system size at $L = 40$, and particle density at $\rho = 2.5$. The $+$ and $-$ symbols denote counterclockwise (CCW) and clockwise (CW) vortical motion, respectively. The different regions marked as $1,2,3,4$ in diagrams. (a) Simulation snapshots at $t = 500$ for a symmetric arrangement of four equidistant regions at noise strengths $\eta = 0.5$, $2.0$ (parallel (top right) and perpendicular (bottom left) configurations), and $5.0$. (b) Simulation snapshot at $t = 500$ for an asymmetric inverted L-shaped arrangement of three regions, using the same noise strengths as in (a). Remaining parameters are provided in Table~\ref{tab:parameters}.}

    \label{fig:fig5}
\end{figure*}

Previously, we established that the geometry-locked flocking observed in the intermediate noise regime ($1<\eta<3$) arises due to periodic boundary conditions, which effectively creates a square lattice type of arrangement of ordered regions. 
To determine whether these emergent modes dynamics can persist within complex environments, we extend our model to a discrete lattice of well-separated non-noisy domains (Fig.~\ref{fig:fig5}(a)). Specifically, we place four non-noisy circular regions symmetrically within the simulation box and ask whether the directional information can be relayed from one region to another. In agreement with our predictions for a single region, the multiple regions exhibit a directional correlation in the geometry-locked state for low-to-intermediate noise range ($1<\eta<3$). Interestingly, the particles split into two flocks of nearly equal density moving either in parallel rows or columns, as shown in Fig.~\ref{fig:fig5}(a). These flocks are connected through the periodic boundary conditions, forming a band of particles moving in the same direction (see Supplementary Movie 5). The stability of the band depends on the particle density. This can be visually seen in the single region case in Fig.~\ref{fig:fig2}(b), where the aligned band becomes unstable and disappears as density is lowered. In contrast to the geometry-locked state, the vortex state randomly selects clockwise or counterclockwise directions (see Supplementary Movie 6). As the presence of a high surrounding noise does not allow the flow of directional information to the neighboring regions. Consequently, we do not observe any correlation in the vortical motion of particles between the regions. This behavior persists even when the size of the regions is increased to narrow the gap between them. In fact, the vortex state weakens because the surrounding noisy region required to sustain the state is reduced. 

We next ask what happens when multiple non-noisy regions are asymmetrically placed within the global environment. To this end, we simulate a configuration of three regions arranged in an inverted-L geometry \cite{Souslov2017} (Fig.~\ref{fig:fig5}(b)). In the geometry-locked flocking regime ($\eta=2$), the particles spontaneously select a single branch to move (see Fig.~\ref{fig:fig5}(b) and (c) and Supplementary Movie 7). Unlike the symmetric case, the flocks do not split into two branches, a result likely driven by the asymmetry of the orthogonal allowed directions of motion. As a consequence, the formed flock is dense and wider than in the case of symmetric arrangements ($\eta = 2$, Fig.~\ref{fig:fig5}(a)). The vortex mode remains uncorrelated and does not show a different dynamics due to the change in the geometrical configuration. Taken together, these results suggest that by tuning the noise and size of the regions, geometry-locked motion can be correlated across multiple regions in space, while the vortex mode remains decoupled.

\section{Inducing anti-ferromagnetic order in the vortex state} 
\begin{figure*}[t]
   \centering
   \includegraphics[width=1\linewidth]{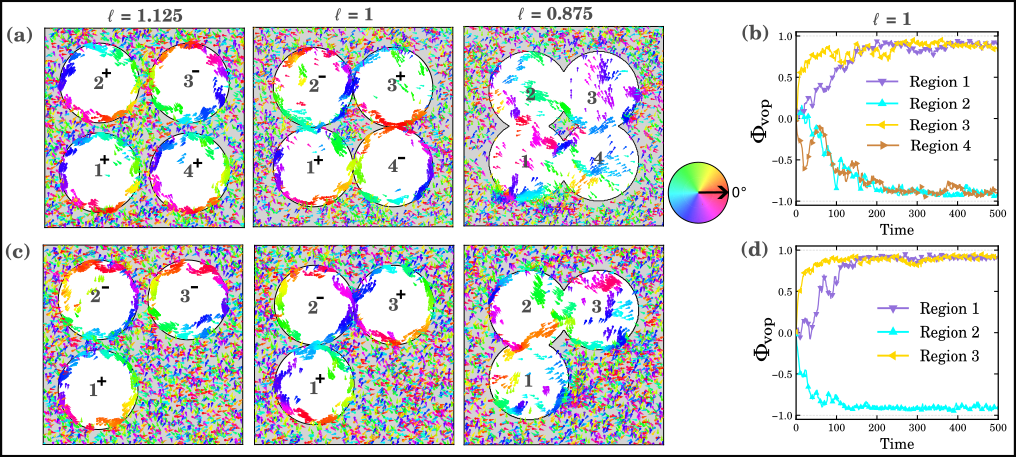}
   \caption{\textbf{Inducing spatial correlation in the emergent vortex state for multiple non-noisy regions.} (a) Directional correlations between non-noisy circular regions in symmetric arrangements as a function of the normalized center-to-center distance, defined as $\ell = d / 2R$. Values $\ell > 1$ and $\ell < 1$ represent finite spacing and overlap between regions, respectively. Simulation results for $\ell = 1.125$, $1.0$, and $0.875$ illustrate the progression from uncorrelated to correlated vortex states. For $\ell = 1$, the regions intersect at a single point along the line connecting their centers. Positive ($+$) and negative ($-$) signs denote counterclockwise (CCW) and clockwise (CW) circular motion, respectively. (b) Vortex order parameter for individual regions in the symmetric arrangement at $\ell = 1$, exhibiting anti-ferromagnetic order. (c) Directional correlation between regions for the asymmetric arrangement at $\ell = 1.125$, $1.0$, and $0.875$. (d) Vortex order parameter for regions in the asymmetric arrangement at $\ell = 1$. The system size is fixed at $L = 40$, and other parameters are provided in Table~\ref{tab:parameters}.}

    \label{fig:fig6}
\end{figure*}

The coupling of active motion in two-dimensional lattices has been previously studied in several experimental works, which demonstrated anti-ferromagnetic coupling under certain conditions, primarily governed by lattice spacing \cite{Han-Sokolov2023, Nishiguchi2018, Han2023, Wioland_ferro2016}.  Our system lacks such correlation because the strong surrounding noise shields information flow from the non-noisy region to the outside environment. Microscopically, the shielding originates from the cancellation of radial velocity components at the interface between the noisy and non-noisy regions, promoting local vortical motion while suppressing information transfer (see Supplementary Fig.~S3). Therefore to sustain a correlated vortical motion one need to strengthen the information flow even in the presence of a strong noisy surroundings. To achieve this, we boost the vortex motion within the non-noisy regions by embedding the lattice within a wider noisy environment while keeping the inter-region spacing small (Fig.~\ref{fig:fig6}). This configuration increases global vortical motion along the edges, which can be sustained by individual circular regions only through antiferromagnetic ordering.

Within this setup, we study the response of the system to variations in the normalized inter-region spacing  ($\ell$) for both symmetric and asymmetric arrangements. As shown in Fig.~\ref{fig:fig6}(a), the dynamics remain essentially uncorrelated under a moderate reduction in center-to-center spacing ($\ell=1.125$). Only when the regions are in close proximity ($\ell \approx 1$), a spatial correlation is observed; diagonally opposite regions rotate in the same direction, whereas neighboring regions rotate in opposite directions, forming an antiferromagnetic vortex order in the lattice (see Supplementary Movie 8). The temporal evolution of these correlations is quantified through the vortex order parameter of each region, as shown in Fig.~\ref{fig:fig6}(b). Upon further reduction of the separation, the four circular domains begin to overlap, destabilizing the vortex state and driving a transition back to collective flocking ($\ell=0.875$, Fig.~\ref{fig:fig6}(a,c)). As the overlap increases and the domains eventually merge into a single circular region, stable circular motion is again recovered. Similar behavior is observed for asymmetric arrangements of non-noisy circular domains, as shown in Fig.~\ref{fig:fig6}(c,d). We further examine a triangular lattice geometry, which constitutes a geometrically frustrated arrangement for antiferromagnetic vortex ordering. In this case, the vortex phase is all together destroyed as the inter-region distance decreases, as strong vortical motion with ferromagnetic order is not feasible (results shown in the Supplementary Information Fig.~S4). Taken together, these results identify the key conditions required for the emergence of correlated vortex motion in large lattices. These results align with recent studies that demonstrate the dependence of inter-region spacing on the formation of antiferromagnetic vortex order in lattices formed by local circular confinements~\cite{Wioland_ferro2016, Han2023, Han-Sokolov2023}.

\section{DISCUSSION}

Living active matter frequently encounters environmental heterogeneities—such as temperature variations, patchy nutrient landscapes, and physical barriers $-$ that cause localized subsystems to favor different ordered states than the bulk. Understanding how these spatial variations shape emergent patterns remains a fundamental challenge, the solution to which could pave the way for manipulating active matter for technological applications. Here, we demonstrate that the Vicsek model consisting of a non-noisy region exhibits two distinct modes of motion: a geometry-locked mode characterized by directed transport and a vortex mode defined by internal vortical motion. These motion modes are tunable via external noise, particle density, and domain size and can correlate across the spatial configuration of the non-noisy regions. 

Mechanistically, these distinct modes of motion arise via different pathways. The geometry-locked mode emerges from a subtle loss of spatial symmetry that does not disrupt the collective behavior but biases the system's directional choice toward the path of maximal directional persistence. This behavior is reminiscent of directed transport reported in other active systems \cite{bricard2013emergence, Xu2024, Wioland_2016,Morin2017}. The vortex state emerges at high surrounding noise without physical boundaries, driven entirely by the sharp noise contrast at the domain interface. This phenomenon stands in contrast to typical vortex states reported in literature, which generally form under circular confinement \cite{Wioland_Hugo2013, Qu2021, Wu2017, Lushi2014, Liu2021, Lin_Guozheng2025}, within attractive systems that aggregates moving particles \cite{Barberis2016, Wang2023, Orsogna_2006, Delcourt2016}, or in systems with inherent chirality \cite{Patra2022, Caprini2024, Dunajova2023, Ventejou_2021} or limited vision models \cite{Saavedra2025}. In this regime, particles are not spatially trapped; instead, their velocity vectors become modulated differently along radial and tangential directions, sustaining stable vortical dynamics. Tracking the velocity components of particles near interface reveals how tangential motion develops spontaneously at the inner boundary over time (Supplementary Fig.~S3). A hydrodynamic theory based on the Toner--Tu framework \cite{toner-tu1995, Marchetti-Ramaswamy-2013, hydrodynamic, TONER-2005} for our setup supports these observations, predicting a transition from a globally ordered nematic order to a vortex order as the regional order parameter contrast between the domains increases (Supplementary Fig.~S5 and Movie 9). Beyond the specific conditions demonstrated thus far, we systematically evaluated the robustness of the emergent patterns in our model. A finite-size scaling analysis reveals that the trends for both $\Phi_{\mathrm{top}}$ and $\Phi_{\mathrm{vop}}$ are remarkably robust against finite-size effects, collapsing onto a single curve upon appropriate scaling (Supplementary Fig.~S6). Furthermore, we varied the spatial gradient of noise across the interface (via a Hill function) and found that the steepness of the noise profile significantly influences the circular order strength (Supplementary Fig.~S7).

A significant advancement of our model compared to previous studies \cite{SpatialNoise, Huang_jun2024, Huang2024, Huang2025, khan_2026} in this domain is the persistence of the emergent states in large systems containing multiple non-noisy regions. Even though both emergent states arise within the same system and represent variations of flocking behavior, their spatial correlations across domains differ fundamentally. While the geometry-locked mode easily propagates directional information between neighboring domains due to the moderate surrounding noise, the vortex mode suppresses this propagation because of the intense surrounding noise. This limitation can be circumvented, however, by introducing a wider region of high surrounding noise around all the regions while keeping the internal gaps small, which ultimately drives the system into a stable antiferromagnetic-like order. These results qualitatively align with experimental observations in active magnetic rollers that form globally correlated states \cite{Han-Sokolov2023} as well as bacterial suspensions in obstacle (pillar) lattices that self-organize into hydrodynamically coupled vortices displaying long-range antiferromagnetic order \cite{Nishiguchi2018, Reinken2020}.

In summary, this work uncovers the physical mechanisms by which environmental heterogeneity modulates both the continuous translation and rotational symmetries governing the flocking motion and its long-range spatial correlations. Unlike the standard Vicsek model, where collective orientation emerges via spontaneous rotational symmetry breaking in two dimensions, the non-noisy domains interact with the periodic boundaries and the surrounding environment to induce explicit symmetry breaking. At a moderately noisy surrounding, the geometric periodicity collapses the continuous manifold of directional choices into a narrow bimodal distribution, leading to a geometry-locked state. On the other hand, strong surrounding noise breaks the spatial translational invariance of the motion, confining the flock within the non-noisy domain and forcing a transition from uniform translation to macroscopic chiral vortex motion. By establishing how environmental landscapes can precisely alter collective dynamics, we demonstrate clear design principles for guiding active matter and provide key insights that advance our fundamental understanding of emergent behavior in non-equilibrium systems.

\section{METHODS}

\section{Quantification of Collective Motion}

To analyze the collective dynamics of the system, we compute a set of order parameters, each designed to capture a distinct aspect of the emerging motion patterns.

(a) The global degree of polar alignment is quantified by the Vicsek order parameter, defined as
\begin{equation}
\Phi = \frac{1}{N} \left|\sum_{i=1}^{N} e^{i\theta_i}\right|,
\end{equation}
where $\theta_i$ denotes the orientation of the velocity of particle $i$. The parameter $\Phi \in [0,1]$ measures the degree of global polar alignment in the system, with $\Phi = 1$ corresponding to perfect alignment and $\Phi = 0$ to a fully disordered state. Starting from random initial positions and orientations, particles progressively align their directions of motion over time, eventually giving rise to the ordered collective state known as flocking, in which all particles move coherently along a common direction selected spontaneously from the available axes of the simulation domain.

(b) To quantify the instantaneous collective direction of motion, we introduce the mean direction, defined as
\begin{equation}
\langle \theta \rangle = \tan^{-1}\!\left(\frac{\sum_{i=1}^{N} \sin \theta_i}
{\sum_{i=1}^{N} \cos \theta_i}\right),
\end{equation}
which is obtained from the angular components of all particle orientations and gives the average direction of collective motion at each instant in time.

(c) To evaluate the strength of the fourfold symmetric motion, we define the tetratic order parameter as~\cite{Hou2020}
\begin{equation}
\Phi_{\mathrm{top}} = \frac{1}{N}\left|\sum_{i=1}^{N} e^{i4\theta_i}\right|,
\end{equation}
where the factor of $4$ in the exponent accounts for the fourfold rotational symmetry of the observed state. The parameter $\Phi_{\mathrm{top}} \in [0,1]$ quantifies the emergence of tetratic orientational order independently of polar or nematic alignment, with $\Phi_{\mathrm{top}} = 1$ corresponding to a perfectly fourfold-symmetric state.

(d) To quantify the deviation of the collective motion from the fourfold symmetric axes, we compute the orientation of the tetratic order, defined as
\begin{equation}
\Delta \theta =\frac{1}{4}\operatorname{arctan2}\!\left(\sum_{i=1}^{N} \sin(4\theta_i),
\sum_{i=1}^{N} \cos(4\theta_i)\right),
\end{equation}
which gives the principal orientation of the tetratic phase, defined modulo $\pi/2$ due to the fourfold symmetry, and measures the degree to which particles deviate from the parallel and perpendicular axes of the domain.

(e) To quantify the degree of the rotational motion, we define the vortex order parameter \cite{Wioland_Hugo2013} as
\begin{equation}
\Phi_{\mathrm{vop}} = \frac{1}{N}\left|\sum_{i=1}^{N}\frac{{v}_i \cdot \hat{{e}}_{\theta,i}
}{|{v}_i|}\right|
\end{equation}
where ${v}_i = (v_{x,i}, v_{y,i})$ is the velocity of particle $i$ and $\hat{{e}}_{\theta,i}$ is the local azimuthal unit vector defined with respect to the vortex centre. The parameter $\Phi_{\mathrm{vop}}$ measures the degree of coherent rotational motion in the system, with $\Phi_{\mathrm{vop}} = 1$ corresponding to a perfect vortex state.

\section{Data and Code Availability}
The code and data are available from the authors upon reasonable request.

\section{Acknowledgements}
The authors acknowledge IIT Kharagpur for providing research facilities and the Government of India for access to the PARAM Shakti HPC facility at IIT Kharagpur. P Patra gratefully acknowledges the financial support from the Faculty Research Scheme for Research Grant (FRSRG), IIT Kharagpur. The authors also thank D. Biswas for valuable comments and suggestions that helped improve the manuscript.

\section{Author Contributions}
P.P., A.S., and M.P. conceived and designed the study. 
A.S. performed the simulations and data analysis. A.S. and P.P. wrote the manuscript. All authors reviewed and approved the final manuscript.

\section{Competing Interests}
The authors declare no competing interests.

\bibliographystyle{cip-v3-bst-submit}
\bibliography{cip-v3-bst-references}



\end{document}